**Aida Slavic**
**UDC Consortium**

**Ronald Siebes**
**Vrije Universiteit, Amsterdam**

**Andrea Scharnhorst**
**Data Archiving and Networked Services (KNAW)**

# Chapter 5
# Publishing a Knowledge Organization System as Linked Data
## The Case of the Universal Decimal Classification[††]

## Abstract
Linked data (LD) technology is hailed as a long-awaited solution in web-based information exchange. Linked Open Data (LOD) bring this to another level by enabling meaningful linking of resources and creating a global, openly accessible knowledge graph. Our case is the Universal Decimal Classification (UDC) and the challenges for a KOS service provider to maintain an LD service. UDC was created during the period 1896-1904 to support systematic organization and information retrieval of a bibliography. When discussing UDC as LD we make a distinction between two types of UDC data or two provenances: UDC source data, and UDC codes as they appear in metadata. To serve the purpose of supplying semantics one has to front-end UDC LD with a service that can parse and interpret complex UDC strings. While the use of UDC is free the publishing and distributing of UDC data is protected by a licence. Publishing of UDC both as LD and as LOD must be provided for within a complex service that would allow open access as well as access through a paywall barrier for different levels of licences. The practical task of publishing the UDC as LOD was informed by the "10Things guidelines." The process includes conceptual parts and technological parts. The transition to a new technology is never a purely mechanical act but is a research endeavour in its own right. The UDC case has shown the importance of cross-domain, interdisciplinary collaboration which needs experts well situated in multiple knowledge domains.

## 1.0 Introduction
Linked data (LD) technology is hailed as a long-awaited solution in web-based information exchange which removes obstacles imposed by platform- and domain-dependent formats and standards. It is also seen as another way to organise information, in graph-like structures rather than in database structures. 'Open' LD, as part of the Linked Open Data (LOD) cloud, brings this to another level by enabling meaningful linking of resources and creating a global, openly accessible knowledge graph with almost unlimited potential for generating new and unexpected associations between dispersed bits of information. The custodians of the LOD cloud, its main technologies and methods, are part of the scientific community of the semantic web. They bring together consortia and forums such as W3C, DCMI (Dublin Core Metadata Initiative), BIBFRAME (Bibliographic Framework Initiative), LD4L (Linked Data for Libraries), semantic web conferences (such as ESWC, ISWC), etc. But, the user base of LOD or LD technology is far broader, encompassing information processes and services in economy, science and society at large. Consequently, the LD technology is a very dynamic and fast-growing field. In this chapter we would like to contribute to the exchange of experiences among those who adopt this technology. Our case is the Universal Decimal Classification (UDC), and we will discuss the challenges for a knowledge organization system (KOS) service provider (in our case the UDC Consortium) to maintain an LD service.

---

[††] Over the years, the UDC linked data project has benefited from expert help by Christophe Guéret who was first to propose the UDC linked data as a lookup service supporting a more complex model of UDC publishing, Chris Overfield for his contribution in setting up RDF stores and putting the service together and Attila Piros for developing the new UDC parser. Finally, this project benefited greatly from the DiKG project for making it possible to bring the UDC LD data project to completion.

The UDC as a showcase provides an insight into the reasoning, procedures and challenges associated with applying LD in the bibliographic domain, specifically with respect to KOSs as bibliographic tools. The expression bibliographic sector or bibliographic domain covers activities, agencies and services concerned with preserving, collecting and organizing recorded information and facilitating information discovery and access since the beginning of literacy. The bibliographic domain comprises the library sector, information and documentation centres, the publishing sector, services such as bibliographic and abstracting services, citation indexes and full text bibliographic services (e.g. Citation Index, Chemical Abstracts, Ebsco, Inspec, ARIBIB). Since the end of the 19th century the bibliographic domain has been creating international standards for describing and indexing information resources, such as cataloguing standards, KOSs, and later, in the computer era, data coding and bibliographic data standards (e.g., the MAchine-Readable Cataloging or MARC family of standards), data and service protocols, etc. These tools and standards have been enabling international exchange and cross-collection information discovery among libraries and between libraries and bibliographic services or publishers.

The UDC was one of the pioneering tools designed to meet the growing information needs of industrialisation and to support opportunities for learning and bettering the society which came with it. To better understand the challenges when publishing a bibliographical tool such as UDC as LD, in Section 2 we will remind the reader of some fundamentals of information and knowledge organization as applied in the bibliographic domain. These are very basic, but often, maybe because they are so basic, they are not articulated or made explicit. We continue in Section 3 to present the UDC as an exemplary case, outlining the main features of a classification and its own trajectory into automatization.  In section 4, we elaborate on the identified challenges when it comes to sensible LD publishing of a system as complex and long lived as the UDC. Section 5 presents the technological choices made to respond to those challenges. In section 6, we conclude by summarizing our understanding of how legacy collections and tools used in information discovery can enrich and inspire the ways in which the LD technology may be applied in the future.

## 2.0 Information organization, knowledge organization and linked data
## 2.1 Visions and foundations
Until the advent of the internet and the semantic web, i.e. LD technology in particular, a bibliographic domain was a relatively isolated information space with limited options for merging with or allowing information linking with other information domains (scientific data sources, archives, musea, etc.). Similarly, KOSs developed in the bibliographic domain such as bibliographic classifications, subject heading systems, thesauri and descriptor systems have only occasionally been used outside their field of provenance. Thus, they could not be used as a link between similar information contents dispersed in different information domains and sectors.

This is rather disheartening given that both the Mundaneum by Paul Otlet in 1910 and the Memex by Vannever Bush in 1945 were envisaged as services to enable easy access to all records of human knowledge (van den Heuvel 2008; Wright 2014). They envisaged a knowledge space where we would seamlessly access and move between primary (documents), secondary (bibliographic data) and tertiary information sources (encyclopaedias). These ambitious and visionary projects firmly rooted in the bibliographic world were destined to fail simply because they were not supported by technology similar to what is at our disposal today. The LOD cloud, as a manifestation of the semantic web, has remarkable resemblance to the visionary ideas of a place where we can access all human knowledge (The Linked Open Data Cloud 2020). If we look at the LOD cloud diagram we can discover many LD clusters, representing both KOS and contents indexed by them in different fields of human activities. We can see bibliographic data clouds, i.e., secondary sources, being connected to KOS clouds and both being linked to primary sources and scientific data. All of these clouds are connected to tertiary information sources: encyclopaedias and other reference sources. It is noteworthy, that the centre of the LOD cloud is occupied by DBpedia, the LD representation of Wikipedia. As it can be observed from the LOD cloud visualization, the semantic web operates in the realm of "everything," "universal," "all" and although bound to the digital world, it does not draw spatial, linguistical or temporal boundaries with regards to information and knowledge. This analogy was the main motivator behind making the UDC one use case in the Digging into Knowledge Graph project (DiIKG; see Martínez-Ávila et al. 2018; Szostak et al. 2018; Szostak et al. 2020).[1]

It seems natural to assume that with the help of LD technology, bibliographic data and tools such as KOSs are on their way to being fully integrated in the semantic web where they are much better placed to fulfil their role of a pivot connecting different knowledge structures and content. However, as we experienced, on the implementation level we have to deal with many details and complexities which have the potential to turn into obstacles. Once we resolve the basic technological obstacles of identifying and linking resources, everything else depends on the semantics, i.e., on the question of whether the premise upon which two things, two concepts or two resources are related is true. Identifying and connecting named entities and information objects in the Web space is relatively easy and straightforward. Connecting ideas and knowledge about these entities and preserving the many ways these may be systematized, represented and communicated in human knowledge is an entirely different plane of complexity. This is why it is important to create a shared understanding when it comes to notions such as "concept," "resource," "value," "label," "term," etc. (cf. Smiraglia and van den Heuvel 2013).

There are many "meta" levels through which recorded knowledge is communicated and there are methods and semantic models developed throughout history from the beginning of literacy. In the domain of recorded information, we manage information and knowledge by differentiating between concepts (thoughts), languages by which we communicate these thoughts, abstract bodies of work in which thoughts are organized (intellectual work), expressions we use to communicate these bodies of work (painting, speech, textbook, fiction), the embodiments of this work into a certain physical format and recording onto some kind of carrier (book, image, recordings, file). We manage information discovery by separating information resources, with all their facets as listed above, from their content (aboutness) and we use metadata to aggregate, present and retrieve information.

## 2.2 Aboutness and knowledge organization systems

The expression "subject of a document," whereby subject represents a summarised body of ideas, is commonly used in information organization to denote aboutness, i.e., the content of an information resource. When populating metadata to describe what a document is about, we perform subject indexing. In doing so, due to the complexity of human knowledge and ambiguities of natural language, we have to use formalized languages, i.e., indexing languages or more broadly knowledge organization systems (KOSs). KOSs are sources of concepts and associated language terms whose meaning and position in the knowledge space is defined through semantic relationships. They tend to represent knowledge as a coherent system with an associated formal vocabulary and syntax. They are external tools for representing knowledge forms that comply with accepted scientific, educational and professional consensus in a given time and in given domains of application. In the bibliographic domain, KOSs are, conceptually, either classifications or alphabetical indexing languages (descriptors, thesauri, subject heading systems). Classification groups concepts according to their similarity into classes or class categories which are all assigned a notation (artificial code) to preserve the logical order and meaningful grouping of concepts. Alphabetical indexing languages use natural language terms to represent concepts and arrange concepts alphabetically. These two types of KOSs support different functions in indexing and retrieval and are often used in combination.

In indexing these KOSs ensure predictability and in information retrieval they can be implemented to support information browsing and semantic search expansion. The strength of KOSs is that they are standalone, self-contained, external resources that can be shared and are often developed or function as standards for the international exchange of information. The owners and curators of KOSs can be standards bodies, consortia, institutions or agencies concerned with subject access to information who take it upon themselves to maintain and keep knowledge structures and associated vocabularies up to date. More complex KOSs are kept in databases in proprietary or, in best case scenarios, in domain-specific data structures and formats (e.g., *MARC 21 Format for Authority Data* 2020) and are made available for use as printed or digital outputs or as web applications. In libraries, at the point of use, they can be integrated with library systems in the form of subject authority files. Far more frequently, indexing terms taken from a KOS will be found only as values in a dedicated field of bibliographic metadata where they appear as simple textual strings detached from the semantic context of the KOS from which they were taken (Slavic 2008). Although many KOSs are generally appreciated as authoritative sources of terminology and useful tools in providing semantic relationships, their use outside the bibliographic domain has always been limited by the lack of their accessibility and common vocabulary exchange standards.

This has all changed with LD and the availability of the Simple Knowledge Organization System (SKOS) standard, OWL (Web Ontology Language) and associated tools and standards for porting KOSs into the semantic web, which gained momentum from 2009 onwards (Slavic 2016). The discussion of exposing KOS as linked data, however, started fairly early in the wake of the semantic web phenomenon (Zeng and Mayr 2019). Simpler controlled vocabularies such as thesauri proved to be more accessible to non-experts and easier to model by the SKOS developers. Thus, thesauri on different subjects and in a range of languages were more readily published as linked data using the SKOS standard. Larger and more complex systems such as general classification schemes traditionally dominating international information exchange within the bibliographic domain have been somewhat lagging behind due to a combination of factors related to their publishing models and limitations of SKOS (Slavic 2016). This is likely to change with the Library Linked Data (LLD) (Baker 2011; Tillet 2017) and BIBFRAME initiatives gaining prominence in the bibliographic domain and with the increasing number of bibliographic metadata collections and bibliographic tools such as name and subject authorities being published as LOD. They are creating a natural environment in which bibliographic classifications are a missing piece of the puzzle. With both bibliographic metadata and KOS data residing in the same space within a LOD cloud, it is possible to connect indexing terms in, for example bibliographic metadata, with the KOS from which these terms were taken and where they have further semantic links or to use classification as a pivot for mapping between KOSs. But this also means that some of the idiosyncrasies, procedures and approaches in supporting subject access, developed and being used in hundreds of thousands of legacy collections, are now also becoming a part of the semantic web story.

KOSs are numerous with different provenances and knowledge structures linked to various fields of application. They predate LD technology and even when kept in a machine-readable data format they are not likely to be modelled with the level of formality typical of ontologies. Thus, every KOS published as linked data has to undergo a process of converting its data model to a readily available but simplified model such as SKOS. This usually means either dumbing down sophisticated semantic features or extending the SKOS data model with elements from other web ontology standards. Further to this, publishing KOSs as LD, and in particular as LOD, is associated with many levels of decision making. In this chapter we will illustrate some of these using the example of the UDC.

## 2.3 Why classification?

Classifications deal with meanings and thoughts as elements of a knowledge space as a whole and, in the case of general knowledge classifications, such as UDC, this means all areas of human activity. The defining feature of classifications is that they deal with concepts, i.e. ideas, and are not concerned with the language used to express them. Classifications group and organize concepts according to their semantic proximity into a logical sequence of classes and subclasses constituting a hierarchy. Each class may comprise any number of concepts sharing the same characteristics. Classes from a classification scheme may be combined to express complex statements about meaning, following a specific set of combination rules. When we use a classification, we can group things that belong together, we can present them in a logical order and we can do so irrespective of whether we describe simple or complex subjects. This makes classifications indispensable when it comes to the logical presentation, organization and contextualization of information within a knowledge space. They can represent what is already known and what is deemed important to communicate for a given purpose in an information environment. By providing statements on how ideas are understood they are not only instructive but also helpful in discovering new things and anomalous phenomena through the analysis of patterns, interpretation and derivation from the existing structures (Kwasnik 1999).

In the bibliographic domain classifications are used to communicate and facilitate discovery and access to knowledge. The structure of these classifications is derived from knowledge as recorded in documents, i.e., as treated in a society, culture, science and education (Smiraglia 2001). Almost all general bibliographic classifications[2] organize knowledge in a series of disciplines and subdisciplines reflecting the way knowledge is taught and used. Classifications strive to be hospitable (expandable) and extensible to accommodate new and emerging knowledge and they would probably be more successful in doing so were it not for demands for stability and resistance to change imposed by the practicalities of resource collection management (see Suchecki et al. 2012, on the evolution of categorial systems). To mitigate structural rigidity, bibliographic classifications deploy various structural features such as facet analysis, analytico-synthetic features, perspective and alternative presentations of knowledge, syndetics (lateral semantic linking), etc. All of these create difficulties in representing, i.e., expressing, these structures using formal ontological languages.

## 3.0 The case of the UDC
### 3.1 Roles and intracacies of bibliographic classifications
In bibliographic and many other knowledge classification schemes, each class is assigned a unique (alpha)numerical symbol termed a "notation,"[3] which is used to represent its particular scope of meaning and the meaning itself is described with plain text, called a "caption."[4] The expression "classification scheme" denotes a given, named and authorised reference tool containing all notations with their corresponding meaning in multiple hierarchies covering all forms of knowledge and rules for a particular knowledge organization system. Schemes can be translated, published and distributed in many world languages (Figure 1-2). In the process of document indexing, one describes subjects by assigning a notation (or their combinations) taken from a classification scheme and recording this notation in subject metadata.

| notation | caption (class description) | |
|---|---|---|
| | English | French |
| **51** | Mathematics | Mathématiques |
| **512** | Algebra | Algèbre |
| **512.7** | Commutative algebra and algebraic geometry | Algèbre commutative et géométrie algébrique |
| **512.73** | Cohomology theory of algebraic varieties and schemes | Théorie de la cohomologie des variétés et des schémas algébriques |
| **512.731** | Classical topology and cohomology theory of complex and real algebraic varieties | Topologie classique et théorie de la cohomologie des variétés algébriques complexes et réelles |

Figure 1. An excerpt from the UDC scheme hierarchy showing captions in English and French.

| 538.9 | Condensed matter physics. Solid state physics | *English* |
|---|---|---|
| | Φυσική συμπτυκνωμένης ύλης. Φυσική στερεάς κατάστασης | *Greek* |
| | Физика конденсированного состояния. Жидкое и твердое состояние | *Russian* |
| | 凝聚态物理、固态物理 | *Chinese* |
| | ঘনীভূত বস্তুর পদার্থবিজ্ঞান। ঘনাবস্থা পদার্থবিজ্ঞান | *Bengali* |
| | संघनित पदार्थ भौतिकी. ठोस पदार्थ भौतिकी | *Hindi* |

Figure 2. Caption of the UDC class 538.9 in six languages and scripts (UDC Summary).

Classification notations, due to their brevity, are a practical way of labelling printed and other physical media to achieve the systematic arrangement of knowledge in a physical space (e.g., library shelves) and they have been used for this purpose for thousands of years. Equally, notations are used in managing and monitoring the acquisition and circulation of physical documents, as well as organizing and presenting collection metadata for the purpose of information browsing and searching. Because of the fact that they provide a language-independent way of expressing subjects, classifications are particularly useful in information organization and discovery in a multilingual environment. They have a long tradition in being used in this way in the bibliographic domain whether linked to institutions (libraries, museums, archives, documentation and information centres) or information services and agencies supporting research and science, and even more broadly, the publishing industry.

Classifications that have been used for a longer period of time in a larger number of information services often gain prominence and become the tools of choice in information exchange as they can help link similar content irrespective of the language, script, provenance, region, type of information resource or time of publishing. Through their widespread use, classification schemes are translated into many languages and thus gain another useful function: they serve as a reliable source of semantically rich terminology (Figure 2). As such they can be used as a basis for creating thesauri or subject headings or directly to support user friendly interfaces for browsing a knowledge space in multiple languages.

An especially important and useful feature is when a general classification scheme treats all forms and fields of knowledge and their shared commonalities as a coherent system.[5.] Such classifications are typically structured according to disciplines respecting the educational and scientific principles according to which knowledge is taught, researched and applied. In a disciplinary organization, knowledge phenomena are placed in the fields of knowledge in which they are studied. This means that many phenomena may occur in several places, i.e., where they are the subject of study, and therefore their full meaning is determined by the context. This is resolved either by linking these concepts across the entire knowledge space, thus creating a network of associative relationships called syndetic structure. If made available to users, these types of semantic relationships can be very useful in resource discovery (Figure 3).

In the following sections, we will document discussions and technological design decisions made in the process of publishing UDC as LD. We will explain the context in which these challenges emerge by providing some information about the UDC's maintenance and use. We, then, present our own approach to solving these challenges. Some of these solutions are of a general nature and applicable to different KOSs.

## 3.2 The origins of UDC in the context of automation

UDC was created during the period 1896-1904 to support systematic organization and information retrieval of a bibliography in the form of a card catalogue called Repertoire Bib liographique Universel (RBU). Paul Otlet and Henri La Fontaine designed RBU to hold the largest record of human knowledge ever assembled to date. The catalogue was to be organized in systematic and logical order using a knowledge classification of great flexibility and detail. They very much liked the solution of presenting knowledge in ten main groups that could be subdivided as required which they found in the *Dewey Decimal Classification (*DDC*)*, but needed a very detailed indexing language capable of expressing complex subjects in science and technology. Otlet and La Fontaine obtained permission from Melvil Dewey to translate and use the basic structure of his classification and went on to develop an analytico-synthetic indexing language with vocabulary that by 1905 already exceeded the size of *DDC* by several orders of magnitude. By 1914, when it was closed down in the wake of World War I, RBU contained over fifteen million entries classified using UDC (Rayward 1990). For over a decade it supported an information service and functioned as a UDC-based "analogue search engine" answering over 1,500 information requests a year (Wright 2014). The RBU project remains an unsurpassed achievement in the field of bibliography and was entered in UNESCO's Memory of the World Register in 2013. Due to its ambition, RBU is sometimes compared to that of the internet and the semantic web (van den Heuvel 2008 and 2011; Wright 2014) and UDC remains its lasting legacy.

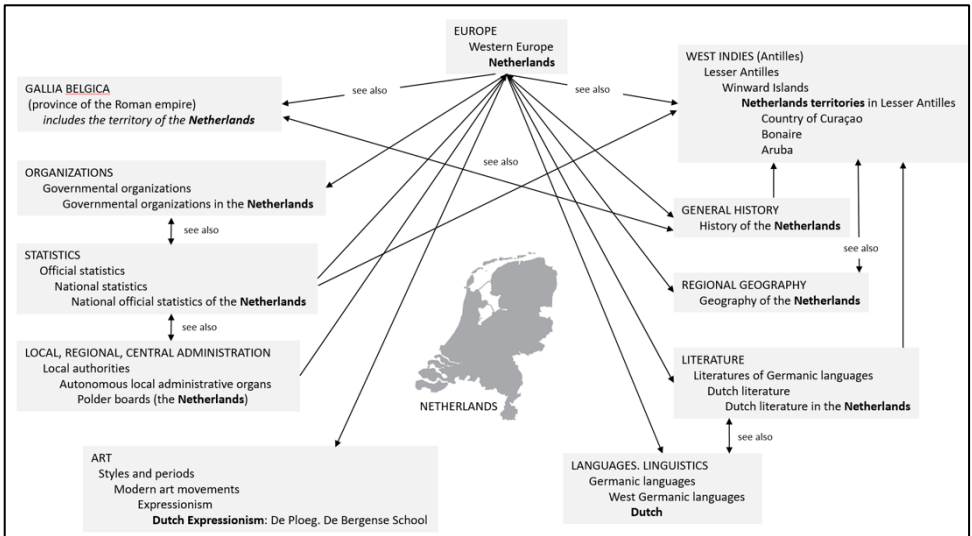

Figure 3. Placement and linking of a concept of "Netherlands" in different parts of UDC.

As a result of the international prominence of the RBU project, UDC has very quickly become the first KOS to be translated, adopted world-wide and maintained in multiple languages. Owing to its size and the amount of detail in the areas of sciences and technology it was often the choice of indexing language for scientific collections and bibliographical databases. As a consequence, UDC was not only the first classification to be used in online information retrieval (from the 1960s), but also the first to be included in information retrieval research, notably Cranfield experiments (cf. Slavic 2005, 14-28). Cranfield experiments measured the performance in the searching of UDC captions and UDC notations (implemented as an analytico-synthetic faceted notation). The UDC notation searching proved superior to other indexing languages in terms of relevance, precision and recall (Cleverdon 1962). The most important takeaway from this research, and the most significant for LD, is that the degree of usefulness of an analytico-synthetic or faceted classification in information retrieval depends on the way it is implemented in the IR system.

There are several other automation milestones reached by UDC that are not as common when it comes to other bibliographic classifications. The English version of UDC was digitized, i.e. stored on magnetic tapes, in the 1980s and converted into a database in 1991, known as the Master Reference File (UDC MRF), containing around 60,000 classes of what has become a UDC standard. Since 1992 the UDC MRF has been distributed to users and publishers as a file export to be ingested and used within an information system or alternatively used within database software provided by the UDC Consortium. With the advent of the Internet there were several projects using UDC either to support an automatic classification of internet resources or to support browsing on information portals and gateways (Slavic 2006). In spring 2001, the standard scheme in English was published on the web, as a service which subsequently evolved into the UDC Hub in nine languages. In 2009, a selection of 2,600 UDC classes was published on the open web as UDC Summary, a database with an online translation interface made available to volunteers, that led to this database being translated into 57 languages.

In 2011, The Multilingual UDC Summary was published as LOD. In the process of publishing this first LD, the UDC Consortium did not envisage any specific purpose these data were supposed to serve. Hence, we refer to this project as an experimental UDC LD phase. Although the need to publish all UDC as LD was clear from the very beginning, requirements and functionalities the UDC LD are supposed to support from the point of view of users and publishers were rather difficult to define due to the factors we will discuss in the following section.

## 4.0 Challenges of publishing UDC as LD

As indicated in the previous sections on the bibliographic domain, we encounter KOSs in two forms: the system itself and KOS terms as they are being used in document metadata. In the same way, when discussing UDC as LD we make a distinction between two types of UDC data or two provenances:

•UDC source data, i.e. the UDC system itself and schedules as they are held in their native database (UDC MRF); and,

•UDC codes as they are applied in resource descriptions and appear in metadata in bibliographical databases, indexing and abstracting services, library catalogues and library shelves.

In the first, experimental phase of publishing UDC as LD from 2011-2019, there were over 2,600 UDC classes made available as SKOS exports. Whilst exposing vocabulary as a SKOS data dump with or without a SPARQL front-end was considered to be a successful completion for many value vocabularies, this was not considered to be the case with UDC.

There are four aspects of the UDC system that require a different approach and special treatment when it comes to publishing UDC as LD:

1. longevity - UDC has been continuously developed and updated for over 120 years;

2. structural complexity - UDC is an analytico-synthetic indexing language;

3. data ownership - UDC is a proprietary system with copyright protected content; and,

4. large usage base and amount of legacy UDC data – UDC is used in document indexing in a variety of bibliographic services, documentation centres and libraries in around 140 countries.

These facts influence the technical solutions that are discussed in this Chapter and deserve a more detailed introduction.

## 4.1 How does the longevity of a system affect LD?

Every KOS has to evolve to accommodate new concepts and terminology in science and other areas of human activity. Although knowledge and associated concepts cannot disappear from an information space or from KOSs, their status can change to obsolete or superseded and old terminology can be declared deprecated and replaced by modern terminology (cf. Tennis and Stuart 2008). Concepts can be moved to different parts of the hierarchy. The fact that the UDC is over 120 years old means that there have been many UDC versions and editions and the UDC system as a whole has a significant amount of historical UDC data and administrative documentation.

Classification schedules comprise a notation (classification codes) and associated text (class caption and notes). The revision of the scheme affects schedules in the following way:

•changes in the class caption and associated terminology affecting the scope and the meaning;
•concepts are moved from one hierarchy to another - in the process the UDC notation is cancelled (deprecated); and,
•new notations, i.e. new classes, are added.

The electronic standard version of UDC, the Master Reference File, created in 1992 comprising 60,000 classes has been undergoing regular change. Subsequent revisions of UDC affected 40% of MRF classes resulting in 12,500 cancelled notations, 22,915 new notations and over 10,000 classes affected by textual changes. The latest MRF contains around 72,000 classes. For all cancelled classes UDC provides replacement data, i.e. redirection to a new valid notation for the same content. A detailed list of changes is distributed to users and publishers with every new MRF release.

These kinds of changes produce a discrepancy between the standard, up-to-date UDC notations and notations appearing in bibliographic databases and libraries world-wide. For instance, up until 1993, UDC notation 94 represented "General Mediaeval and Modern History" and notation 930.9 was "General History. World History (chronological summation of facts)." In 1994, as a result of the UDC revision of the history class (UDC release vMRF94), notation 930.9 was cancelled and replaced by notation 94 which now has a changed description "General History." However, decades after this change there still may be bibliographies and library data world-wide using 930.9 to collocate documents on general history on what is now a non-existing UDC notation.

This is a well-known issue for all users of well-established and widely used KOSs causing permanent tension between requests from users for KOSs to be continuously updated and subsequent rejections of bibliographic agencies to introduce changes in their metadata due to a lack of resources. If both a) bibliographic databases (containing UDC codes) and b) the latest UDC version appear as an LD cloud it might not be possible to establish the link between deprecated UDC codes and those in the most recent version of the scheme. For this reason, if UDC is to serve its purpose in information exchange, it is extremely important to expose not only the most recent version of UDC but also the historical data. This affects the way we model, select the RDF schema and expose UDC data as LD.

## 4.2 How does structural complexity affect linked data?

UDC is an analytico-synthetic and faceted classification which enables the combination of concepts from different areas of knowledge. This feature is very important for the detailed indexing of documents and providing multiple subject access points. The main advantage of this kind of classification lies in its power to express detail and subject range with a relatively small vocabulary. For instance, even using the UDC Summary, which contains only 3,000 classes (out of 72,000 of the complete UDC MRF), we can express the following content:

"Digital audio recordings, in mp3 format, of an anthology of the short stories of the modern Dutch literature of Suriname, in the English language":

821.124.5`06-32(883)(082)(086.7)(0.034MP3)=111

The meaning of notational elements:

821.124.5 Dutch literature (main class for literature);
`06 Modern (special auxiliary for periods);
-32 Fiction/stories (special auxiliary for literary forms and genres);
(883) Suriname (common auxiliary of place);
(082) Anthology (common auxiliary of form);
(086) Audio recordings (common auxiliary of form);
(0.034MP3) Digital documents - mp3 (common auxiliary of form);and,
=111 Documents in the English language (common auxiliary of language).

In order to function in this way, the UDC system consists of a vocabulary (classes of concepts) and syntax rules for combining classes into complex subject statements (McIlwaine 2007; Slavic and Davies 2017).[6] In practical terms, synthesis is enabled by the organization of concepts into tables (facets) and by the notational system, i.e. systems of numerical codes and syntax symbols enabling (de)composition of UDC strings. General concepts can be freely combined with themselves and with all areas of UDC, including the auxiliary tables. As a consequence, generally applicable concepts are always preceded and terminated by a certain symbol or combinations of symbols, punctuation or digits, collectively known as facet indicators, and they are all presented and used in this way throughout the schedules:

=...     Common auxiliaries of language
(0...)     Common auxiliaries of form
(1/9)     Common auxiliaries of place
(=...)     Common auxiliaries of human ancestry, ethnic grouping and nationality
"..."     Common auxiliaries of time
-02     Common auxiliaries of properties
-03     Common auxiliaries of materials
-04     Common auxiliaries of processes
-05     Common auxiliaries of persons

Thus, parentheses (followed by any digit from 1 to 9) always represent place, e.g. (492) represents Netherlands. Language will always be expressed with a number preceded by an equal sign =, e.g. =112.5 Dutch. The main table contains the main classes of disciplines, subdisciplines and fields of knowledge and these classes have a simple numerical notation (used decimally). They can be further specified through combinations with over 15,000 common auxiliary concepts (in the tables listed above), as well as a series of specialized concepts presented in special auxiliary tables that are always preceded either by -, .0 or ` (backtick or inverted comma). All UDC notations from the main tables and those from common auxiliaries can be combined among themselves using the following combination signs/symbols:

+     Coordination
:     Simple relation
::     Order-fixing
[ ]     Subgrouping

Each UDC notation, whether from the main or auxiliary tables can be specified further with alphabetical extensions, e.g. (492Delft) is used to express the Dutch city of Delft. Equally each notation can be connected using * (asterisk) for codes from other systems.

All these features make the UDC an analytico-synthetic system proper which with a relatively small number of classes can produce a very specific description of content (cf. Slavic and Davies 2017).

For instance, countries are listed only once in UDC: in a table of Common auxiliaries of place (place facet) where each country is assigned a unique notation and where the Netherlands is assigned notation (492). Disciplines of geography and history, for instance, are studies closely related to the notion of place and many classifications would need to list the geography of all countries and then the history of all countries of the world. In UDC, however, classes of 913 "Regional geography" or 94 "General history" are virtually empty, i.e., they do not need to list places of the world. Instead, in the process of indexing one combines a notation from the main table 94 and notation from the place table (492) to express 94(492) "History of the Netherlands" or 913(492) "Geography of Netherlands." This can be further extended by adding time auxiliaries, ethnicity, etc. Furthermore, if there is document content describing the relationship between regional geography and the history of the Netherlands, this can also be easily expressed with a combination 913:94(492).

The UDC's analytico-synthetic feature represented by this expressive notational system leads to complex UDC strings of various length (Smiraglia et al. 2013). Each of these strings reveals a precise and rich meaning which can be extracted from the string by accurate parsing of the UDC compound number by both humans and computers. Composition and decomposition of UDC numbers enables easier management and coordination of natural language terms when these are used instead of classification codes in the process of searching (Riesthuis 2008). But, it is also clear that for machine consumption a proper parsing of the string is of utmost importance.

For all the advantages of a partial or fully analytico-synthetic KOS such as UDC there is one important downside when it comes to LD: the UDC namespace does not contain the complex UDC strings that are created in the process of indexing locally and which may appear in bibliographic metadata in many collections world-wide. If UDC is to serve the purpose of supplying semantics, enriching and linking millions of bibliographic records one has to front-end UDC LD with a service that can parse and interpret complex UDC strings and provide adequate resolution by linking each element from the complex string to an appropriate UDC data record. Further in the text we refer to this solution as a lookup service.

### 4.3 How does classification ownership affect the model of LD publishing?

UDC is owned and managed by the UDC Consortium which is an association of publishers and users that operate on a non-profit, self-funded basis. UDC publishers are themselves non-commercial, publicly funded institutions such as national standard institutes and national libraries. The main source of income that sustains UDC maintenance and development comes from the sale of publishing licences or various languages and the sale of UDC schedules. Thus, while the use of UDC is free—the publishing and distributing of UDC data are protected by a licence with separate licences being issued for publishing of up to 50% and for more than 50% of UDC MRF. Clearly, although the business model is non-profit, it is impossible for the Consortium to release the complete UDC schedules as LOD without jeopardizing the future of the UDC.

In order to mitigate this situation, in 2007, the Consortium released, on the open web (under a Creative Commons licence), the UDC Summary, as previously mentioned. This is a set of over 2,600 classes with captions translated in 57 languages. In 2011, this set was published as LOD.

In this context it should be mentioned that the complete content of the up-to-date UDC MRF is available on the web (www.udc-hub.com). Nine of the languages are available through the Consortium's UDC Hub service and two languages as a part of national services (Slovenia and Hungary). For six of these languages, national libraries are paying for a publishing licence to make UDC available free of charge in their respective countries (Croatia, Czech Republic, Hungary, Poland, Serbia and Slovakia). In these countries all the information agencies publishing metadata containing UDC as LD should be able to access and point to the UDC namespace. As time goes by, UDC in other languages is expected to be published in a similar way and free access in these countries will be regulated by publishing licences. Therefore, publishing of UDC both as LD and as LOD are options that have to be provided for, albeit within a more complex linked data service, that would allow open access as well as access through a paywall barrier for two different level of licences.

## 4.4 How does a large user base affect the linked data publishing model?

For over a hundred years UDC has been used in bibliographic databases, documentation centres, libraries and national bibliographies in all parts of the world. There are millions of bibliographic records containing UDC codes that may eventually be published as LOD. Due to its long history and wide-spread use, UDC functions as a *de facto* standard in information exchange. This status is closely linked to the authority and quality control enabled by the stability of its ownership.

Document content analysis and indexing using classification is a costly process and over the past decades, information services in general have fewer resources available to manage proper subject authority data. Automatic classification and indexing, where possible, are not always readily available or adequate and are associated with initial costs. But most importantly, information services have so far had difficulty in passing the benefits of knowledge access on to end users due to poor user interfaces or lack of expertise to exploit the classification data.

In general, classifications are expected to be implemented in an IR system "behind the scenes." They are supposed to support information browsing and searching without users being aware of the complexity and technicalities of an indexing language which is maintained in the background (using authority control). It is incumbent on the classification publishers to provide adequate support to bibliographic agencies and make it easier for them to keep their authorities up to date. This could support the following solution for the benefits of bibliographic agencies and their users:

   •validating and updating classification data held locally (authority files), mapping deprecated notations to new notations or entirely bypassing and converting obsolete classification data in information exchange;
   •enriching bibliographic metadata with additional semantics and verbal access points (additional search terminology in multiple languages, semantic expansion to broader, narrower or related subjects);
   •enabling knowledge graph-based visualization, linear or multi-dimensional presentation for browsing and navigation across collections; and,
   •enabling linking, i.e., mapping, to other KOSs and beyond, where classification acts as a pivot and enables cross-collection subject access.

It is, therefore, logical for the UDC owner to consider exposing the classification as LD not only as an experiment with limited value outside the semantic web community but as a fully functional terminological service which would take on board specificities of UDC data, including legacy data and the way it has been applied in bibliographic collections.

## 5.0 Steps in publishing UDC as LD and LOD

Based on previous experience of publishing the UDC Summary as LD from 2011-2019 and taking on board the challenges outlined in the previous section, this new LD project was significantly larger and more complex and took a longer period of preparation. Thorough planning was especially critical given the objective of publishing different and larger UDC datasets under different access conditions. The steps in publishing UDC as LD/LOD, described in this section are to a large extent informed and follow the procedure described by Siebes et al. (2019).

## 5.1 UDC Summary as LOD 2011-2019: lessons learned

The first experience of publishing the UDC Summary as LD in 2011, proved an important learning step. At the time SKOS was embraced enthusiastically by the community sharing simpler controlled vocabularies such as thesauri and subject heading systems (SKOS 2009). The SKOS data model itself was designed with this kind of KOS in mind: it was simple, lightweight and easy to use and it represented a much-needed vehicle for many simple KOS systems to be published as LOD. SKOS filled in a void within cross-sector standard formats for publishing and sharing controlled vocabularies. It was also readily adopted as a data model in vocabulary management applications in enterprises and the commercial sector to support information and content management. Both KOS and the Semantic Web communities felt an urgency to expose KOSs as LOD, to secure visibility, longevity and find new purpose for traditional quality KOSs. The majority of these vocabularies have been underfunded and in danger of being superseded by advanced text retrieval technology and automatic indexing. This was, especially, the case with KOSs developed by heritage institutions, funded by the public or in the public domain.

The UDC Consortium, which, at the time, considered only a small set of data to be published as LOD (albeit in 57 languages) did not envisage any specific practical use scenario. Therefore, there were only four issues of concern:

- How to map the UDC data model into a SKOS schema?
- What to do with synthetic UDC notations?
- How to select the URI syntax?
- How to publish: as data dump or SPARQL front end?

The mapping of the UDC data structure into a SKOS schema (designed for thesauri) required a bit of tweaking. The general principle for this project was to select the minimum set of UDC data elements, leaving out all administrative and UDC data management fields. SKOS deals with concepts uniquely represented by controlled lexical terms (descriptors). Classification deals with classes of concepts and has three ways for representing and identifying classes: a) unique ID of a class (within UDC database system), b) notation, and c) caption. The most important elements are expressed as follows: the unique UDC class identifier (in the UDC MRF database) was stored as *skos:Concept*, the caption was mapped to *skos:prefLabel* and the notation was mapped to *skos:notation*. The SKOS schema was then extended by several data elements to accommodate UDC-specific notes to differentiate application notes and scope notes. After some deliberation and discussion, it was decided not to extend the SKOS schema to express the difference between simple and complex UDC notations (analytico-synthetic feature). This was left for consideration for the next version of the UDC LD (Isaac and Slavic 2009; Slavic and Isaac 2009). The longest and most protracted discussion took place in relation to URI selection. The dilemma was whether to form a URI containing the UDC notation (the meaning of which depends on the UDC version) which is understandable to humans or whether to use a unique identifier from the UDC database. The final decision, in 2011, was to opt for a numerical identifier thus forming the following type of URI: "udcdata.info/019930" whereby the number "019930" represented a UDC record identifier for the notation 331 "Labour. Employment. Work." An important reason for not including, at the time, UDC notation in the URI was the practice of the occasional re-use of deprecated notations (usually after 10 or more years) which can be a source of ambiguity unless linked to the UDC version. Additionally, UDC notations may contain symbols and signs that according to URL standards require encoding before transmission. Another argument in favour of this decision was that the URI was meant for programs (not people), hence whether or not it contains a notation that can be read by humans makes no difference. The decision to publish UDC LOD as a data-dump and without programmatic access, i.e. SPARQL endpoint was based, primarily, on the fact that the Consortium, at the time, did not have any real life use-case scenario for UDC LOD.

Starting from 2011, the UDC Summary LOD data were available at https://udcdata.info for nine years and frequently downloaded (Figure 4). As we observed, this was mainly for the purpose of harvesting, extracting multilingual terminology and recreating and republishing UDC schedules in local systems or generating new types of proprietary vocabularies under different names. In principle, much of the LD use at the time did not consist of linking within the LOD cloud but rather of downloading, storing and processing data in local systems.

Figure 4. UDC Summary Linked Data (2011-2019) showing UDC class 311, HTML display and its RDF record.

Gradually, following the Library Linked Data (LLD) initiative, more and more library catalogues and national bibliographies became available as LOD and the importance of legacy bibliographic data and the potential use of UDC in accessing and exposing content of these collections has become more obvious. At the same time it was possible to observe and identify potential impediments to linking a UDC standard dataset published as LD with bibliographic metadata sets containing UDC notations (Slavic et al. 2013; Slavic 2017).

LLD clouds observed from 2012 onward consisted of bibliographic metadata (e.g., catalogues of National Szechenyi Library and Library of the Norwegian University of Science and Technology-NTNU) containing very detailed and specific UDC notations. These detailed notations could only be found in the complete UDC standard and were not available within the UDC Summary LOD set. Additionally, much of UDC data in the bibliographic records consisted of complex UDC notational strings which, even if all UDC notations were available as LOD, could not have been linked easily (Figure 5). The main problem identified in almost all LLD sets, however, was the fact that libraries continue to use deprecated UDC notations that are no longer part of the standard active UDC dataset. In some instances, UDC notations found in active library records were cancelled thirty years ago. This means that a UDC namespace, if it is to serve the needs of the bibliographic domain, has to include historical data and concordances between cancelled and new notations.

```
<dc:subject rdf:about="#NTUB00002">
 <rdf:type rdf:resource="http://www.w3.org/2004/02/skos/core#Concept"/>
 <skos:prefLabel xml:lang="no">Abelske varianter</skos:prefLabel>
 <dcterms:udc>512.742</dcterms:udc>
</dc:subject>

<dc:subject rdf:about="#NTUB17121">                          complex notation
 <rdf:type rdf:resource="http://www.w3.org/2004/02/skos/core#Concept"/>
 <skos:prefLabel xml:lang="no">Marine sopper</skos:prefLabel>
 <dcterms:udc>582.28(26)</dcterms:udc>
</dc:subject>

<dc:subject rdf:about="#NTUB00005">                          not in MRF
 <rdf:type rdf:resource="http://www.w3.org/2004/02/skos/core#Concept"/>
 <skos:prefLabel xml:lang="no">Abrasiv slitasje</skos:prefLabel>
 <dcterms:udc>620.178.162.44</dcterms:udc>
</dc:subject>
```

Figure 5. Examples of UDC notations from library linked data (catalogue NTNU).

## 5.2 New approach to publishing UDC as LD

As is evident from our observations, the UDC data found in LLD clouds consists of unstructured, simple textual strings of UDC notations, with no additional data of provenance, semantics or versioning. It is possible that the situation may be the same in local information systems from which these data were derived (cf. Slavic 2008). This means that queries launched from the LLD space to the UDC namespace will contain notational strings only. Thus, any interpretation of UDC data can only come from the UDC namespace, which needs to provide a solution for the linking and semantic alignment between UDC notations in bibliographic records and those in the UDC LD cloud. To do so all problems identified with respect to LLD so far, and as explained in the previous section, have to be addressed.

Clearly, one has to change and improve the way in which UDC LD are published and move from the simple UDC RDF repository to an LD service. This involves not only a change to the amount of data being exposed but the data format and the way these data are accessed. A new approach to publishing UDC as LD requires rethinking the URI format, the RDF schema, but most importantly, instead of a UDC LD-dump or enabling API access, we had to opt for a more complex UDC LD look-up service. In planning this service, we took on board the aforementioned requirements and designed service components to handle the necessary functions.

The basic requirements of the UDC LD service are that, while it has to support the practical use of the scheme in the context of LLD, the service also has to protect the UDC model of publishing and secure its sustainability. The only way to handle this is having a small part of the UDC as LOD and the rest of the licence-protected UDC content available as LD behind a "paywall." In order to resolve problems observed in bibliographic data, the UDC LD-based service has to support the following:

1. Programmatic access to three sets of data:
   - single LOD set: the UDC Summary containing 3,000 classes (under CC BY-NC-ND 2.0 licence)
   - two LD sets behind a UDC MRF licence barrier:
     - Abridged edition (12,000 classes)
     - UDC MRF (72,000 classes), including all twenty versions of the UDC MRF and historical data comprising over 13,000 cancelled (deprecated) classes and their redirections to new classes;
2. UDC Look-up service that:
   - parses and resolves (interprets) a classmark originated from bibliographic data and links its components to relevant records in the RDF data store;
   - upon request supplies URI(s) for UDC notations or the full RDF records.

To meet these requirements the service has to have a more complex architecture and the following solution was chosen:

- RDF stores (three Virtuoso databases: the UDC Summary, the Abridged edition, and the UDC MRF) with SPARQL endpoints accessible only via a restricted RESTful API layer which uses pre-designed SPARQL templates for query execution;
- Web server and custom written UDC parser. The Authentication process is handled by standard shared and private authentication keys (tokens). The HTTP/Get parameters and the HTTP headers inform the server about the type of desired result (e.g., HTML, RDF-Turtle, JSON).

The architecture diagram in Figure 6 shows the UDC LD infrastructure and data flows that support the UDC notation (classmark) lookup process. The service is accessed by initially acquiring an authentication token from the authentication layer. The token received allows access to one of the following services: UDC Summary (LOD), complete MRF data (LD) or Abridged UDC LD.

For parsing and resolving a UDC notation the lookup service receives a plain text URI-encoded UDC notation query e.g. 94(492) encoded as 94%28492%29. The full-service query for, e.g., the UDC Summary, would look like this http://udcsummary.ud-cdata.info/api/parse/94%28492%29. The REST-API receives this query and extracts the URI encoded notation, passes this notation to the Parser which breaks down any compound, synthesized notation into constituent elements. The REST-API receives this from the Parser and retrieves UDC encoded URIs (see section 5.3.3) for each simple notation and returning the results in the required format. The user can then request the full record from the service using the returned URIs.

Other features of the UDC Look-up include an HTML interface for human interaction with the service (Figure 7). The assumption is that the API would be queried by programs submitting simple or complex UDC notations either to get correct URIs for UDC codes or to retrieve complete RDF records. In this scenario, the HTML interface allows humans to verify and explore the provided notations in which the parse tree, versions, and RDF translations are expressed.

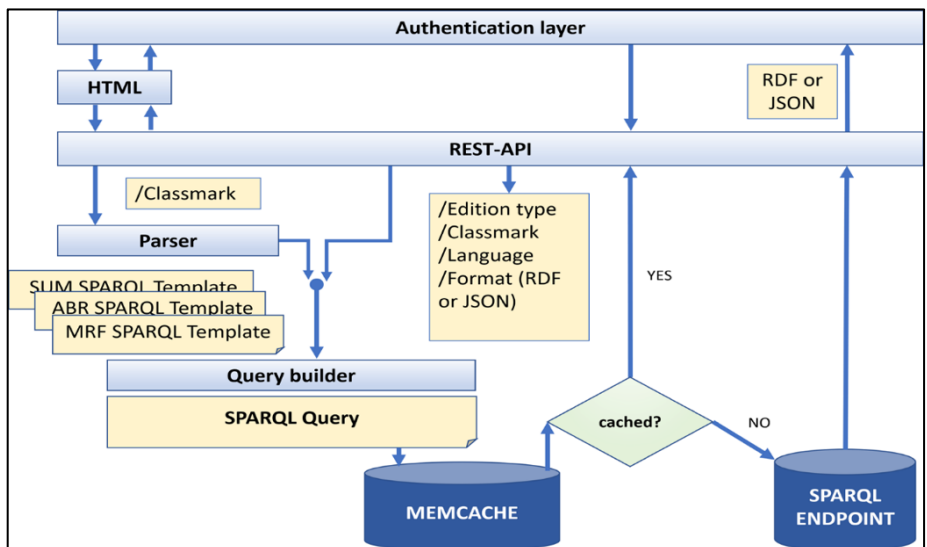

Figure 6. UDC Look-up service architecture.

### 5.2.1 UDC notation parser: an important component of the UDC look-up service

Within an information system, typically a library system, UDC is often managed within a subject authority file which allows managing and searching notational components and associating them with lexical terms and relevant semantic data. An authority control separates and detaches classification data from the way this is used or displayed on a searching or browsing interface. When bibliographic records are exported from a library system, in the process of information exchange or when published as LD, they show UDC notational strings, e.g., as we can see above in the NTNU RDF record 582.28(26). In order to be able to link components of complex UDC strings to their record in the UDC namespace, one has to be able to parse complex UDC notational strings. This function is one of the most important components of the UDC look-up service.

Programs for automatic UDC notation parsing were developed in the past, and as reported by Riesthuis (1997, 1998), his own program was 100% successful in splitting complex UDC strings into searchable components. This is due to the fact that UDC has an expressive notational system and uses facet indicators consisting of punctuation symbols and digits, to indicate the beginning and the termination of a notation for specific types of concepts (as explained earlier in Section 4.2). The absence of punctuation in connection to numbers is also meaningful. Thus, in the following UDC strings 94(492), 94:33 or 94"19"(492) we can clearly see that these are complex expressions consisting in the first two cases of two and in the third case three separate notational elements.

The UDC syntax rules determine which notations can be combined with which other notations and in which linear order. These rules are formalized through the use of selected digits, punctuation and characters and represent a UDC formal language. As UDC develops, over time, some syntax rules get refined or changed, and new notations and notational symbols are added. For instance, since Riesthuis wrote his programme, in 1996, two new common auxiliaries for properties and processes were added. This means that these three characters -02 now denote the beginning of concepts from the table of properties and -04 denotes concepts of processes. Clearly, over time, new algorithms have to be added on top of the previous set of parsing algorithms. For this reason, and for the purpose of the LD Look-up service, Attila Piros has developed a new and improved parsing program that will allow for the continuous and controlled update of different sets of parsing algorithms. This was based on his previous parser known as Piros UDC-interpreter (Piros 2015 and 2017).

The UDC notation interpreter, as put by Piros, is an automata that, based on a series of algorithms, recognizes this formal UDC language and generates a tree which contains the parts of the notation (based on predefined rules) as well as connecting symbols. The basic set of parsing algorithms deals with connecting signs and notations for common auxiliary tables (general concepts) as these are the most stable part of the UDC syntax rules. This is followed by a series of smaller parsers handling subsequent rules pertaining to different parts of the UDC schedule. The very last phase of the parsing process deals with semantic analysis and UDC notations created through parallel division and application of special auxiliaries. Figure 7 shows an HTML interface in which a complex UDC number is split into components that, in this case, have valid UDC classes. The service executes queries against the UDC Summary, the UDC Abridged edition or the UDC MRF and in the second step it generates an RDF representation of the information selected by the user or program.

Figure 7. UDC Look-up service and interpreter.

Prior to arriving at this solution there were several important steps and key decision to be made that are relevant for many linked data projects (cf. Siebes et al. 2019).

## 5.3 Important steps and decisions in publishing UDC as LD
### 5.3.1 Selection of data
In the previous section we mentioned three different UDC datasets: UDC Summary, UDC Abridged Edition and UDC MRF (the complete UDC dataset). The main purpose of the UDC LD service, at this point in time, is to provide support in interpreting and linking content of legacy collections. In doing so, one has to balance constraints of data ownership, licensed users and related context that would enable UDC to sustain. The UDC Summary, the UDC Abridged Edition, and the UDC MRF are maintained in different MySQL databases and the same set-up is replicated for the RDF store (three Virtuoso databases). The selection of these three datasets is based on the well-established practice in UDC data use and publishing. They are representative of two kind of access to UDC data: open access and access through a UDC MRF license requiring an authentication process based on authentication tokens (managed outside the service itself). With respect to the supported languages, the UDC Summary contains language data in 57 languages. Abridged and MRF translation databases contain 14 and 13 languages respectively. The UDC LD look-up service will be incrementally developed to include languages for which the UDC Consortium has a license clearance. In the first phase the Abridged and MRF datasets are planned to be exposed in English only. UDC data comprise many data elements that are required for data management and publishing, for the LD, we selected only 14 data elements (see 5.3.5 below). In terms of sequence of data release, the UDC Summary (the LOD set) was given priority due to the large community of users.

When it comes to the selection of UDC data, the most important and innovative aspect of the UDC LD service is that the UDC MRF dataset includes over 13,000 cancelled, non-active UDC notations. These are the result of the classification revision from 1992-2018. Further cancelled historical records will gradually be added through the process of digitization of the old UDC editions from 1905 to 1992. Once this process is completed the UDC MRF will capture the dynamics and changes in its knowledge structure from the beginning of the twentieth century to date.

To understand the significance of historical UDC data for legacy collections one has to be aware of the nature of changes that affect classifications of knowledge and notation lifecycles. Once concepts and subjects become part of human knowledge they do not disappear, but as the understanding of a field of knowledge changes this affects the concept organization of the field. A good illustration is the classification of plants and animals that evolves based on new knowledge, whereby living organisms are re-grouped in a different way. In the process, an organism can be moved to a new class of organisms and therefore is assigned a new UDC notation. The old notation for that organism is cancelled. From the point of view of UDC data maintenance, the record of the cancelled notation is marked as "not-active" and information regarding replacement notations, i.e., the new notation to which this concept was moved, is supplied.

There are millions of bibliographic records containing UDC notations that were cancelled and replaced by new notations decades ago. The possibility to query historical data, either to retrieve URI for these deprecated notations or to use these data to find redirections/replacement for these notations is very important for libraries, bibliographic services and legacy collections in general. Without this link between an old class and the new class, provided in the UDC namespace, we would not be able to establish meaningful connection between information resources dealing with the same entity in different points in time.

### 5.3.2 Use scenarios, serialization and resolution of UDC codes and URIs

We considered various scenarios in which the UDC namespace would likely be accessed in order to select an appropriate LD serialization of the UDC data source. The service is primarily aimed for access by programs and for this to work there is a need to have disambiguation mechanisms and clear guidance for programmers with respect to various choices that would apply. For instance, often the only information libraries have about the UDC are the classmarks and the location of the UDC Look-up service. Libraries launching queries are not likely to be aware of the UDC MRF versions, including whether notations contain valid or cancelled (deprecated) UDC notations or whether they have licence, i.e., authentication token, to query full UDC data. Their queries may have the following format "udcdata.info/582.281.1(035)." The UDC Look-up service will parse and resolve such a query returning information that notation 582.281.1 is cancelled and replaced by 582.244 and might also return an RDF statement with sets of URIs expressing the relationship between these two numbers. If subsequently a program or person, without an access token for this dataset tries to query these URIs at the UDC namespace, the authentication layer would prevent this request from being executed and return a meaningful error message, combined with some information about the result of the query (e.g. a broader class shared both by the MRF version and the UDC Summary version, i.e. the UDC LOD set).

### 5.3.3 Defining the URI naming strategy

The main principle of the URI is to be durable and well structured. When it comes to UDC LD, the URI strategy involves:

- the choice of the internet domain name;
- the choice of the structure of the URI with respect to sub-domains, notation and datasets; and,
- the solution for issues caused by the URI encoding standard.

In making these decisions we took on board the architecture of the RDF store, i.e. number of separate datasets (including constraints on their access) and the ways the service will be used/queried and the use scenarios.

**Domain name**. The UDC LD domain name "udcdata.info" was established in 2011 and remains the same. The old LD RDF store and LD data dump were taken down in 2019 upon the release of the new UDC MRF12 version and to avoid ambiguity. Given that the look-up service will operate on three different datasets (one of which will be LOD; two are behind the data barrier and would require an authentication token. We added three subdomains in the following way:

**udcsummary**.udcdata.info/...
**mrf**.udcdata.info/...
**abridged**.udcdata.info/...

**URI paths**: As explained previously in Section 5.1, the selection of URI in the first UDC LD version back in 2011 took much deliberation. The choice, in the end, was to use an unstructured type of URI which does not contain UDC notation and the URI for class 311 was based on the database identifier and appeared as follows:

http://udcdata.info/018809

whereby 018809 represents a unique identifier in the UDC MRF database for class 311. Based on the projected use scenarios, we have established that the notation is the only element on which this kind of service can operate and it makes sense that it forms a part of the URI. This decision, however comes with the complications which follow.

**Notation in UDC** is not a unique identifier: Due to the size and longevity of the system a number of notations were cancelled in the past and re-used with a different meaning. Because of this, the meaning of a notation is determined by the UDC MRF version in which it is introduced.

**Which MRF version was used in the URI**?: Once introduced a notation will appear in many subsequent UDC MRF versions. However, the URI is formed from the name of the MRF version in which this class was first introduced. These data are controlled through the Introduction Date field in the MRF database. The URI for class (492) will, thus, indicate that this class was first introduced in the UDC MRF release version coded as 'mrf93'.

udcsummary.udcdata.info/mrf93/(492)

**The issue with URI encoding**: UDC notations contain signs and symbols and these would be URI encoded automatically in the process of making an HTTP request. Thus, a URI "udcsummary.udcdata.info/mrf93/(492)" would appear encoded as "udcsummary.udcdata.info/mrf93/%28492%29." To have better control over the way UDC notations are displayed the service has its own URI encoding. For instance, parentheses (round brackets) will be replaced as follows: opening parenthesis will be "_or_" and closing parenthesis will be represented as "_cr_."

The encoded URI for class (492) in the UDC Summary LOD is as follows: "http://udcsummary.udcdata.info/mrf93/_or_492_cr_" (Figure 8). A UDC look-up service will transpose conventional notations into a required URI format. Therefore, one can query the service using a conventional UDC notation and will receive a rrelevant URI.

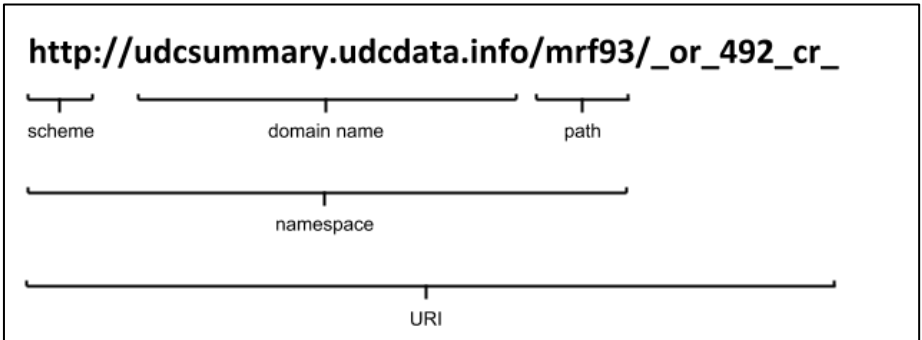

Figure 8. Example of a UDC look-up service URI.

This change of the URI format, as described above, means that a new service must contain a mapping between the old 2011-2019 URIs and the new URI systems.

### 5.3.4 UDC data analysis and RDF presentation schema

UDC MRF is held in a database which contains, apart from the basic UDC data, many administrative data elements relevant for system management, maintenance and publishing. Many of them deal with identification and tracking of changes and continuous revisions to the system. Rather than mimicking the MRF database structure the RDF representation and associated UDC knowledge graph capture the most relevant elements of the UDC system structure and their relationships: class identification data (identifiers, date of introduction, date of cancellation), notation, caption (lexical data associated with language attributes) and semantics. A UDC class modelled as an RDF triple is shown in Figure 9.

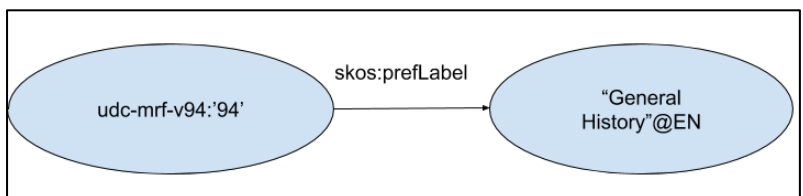

Figure 9. RDF graph representation of class 94 and its caption General History.

Figure 10, below shows a snapshot from the current web-interface of a top level UDC scheme. Next to it we show the corresponding UDC knowledge graph for class 94 General history with its broader class 9 Geography. Biography. History. Both classes were introduced in UDC MRF version v94 and "udc-mrf-v94" combined with notation 94 and 9 respectively, uniquely identifies these two classes.

Figure 10. Top level of UDC structure in UDC Summary with corresponding RDF graph representation of class 94.

The UDC MRF contains both simple and combined, i.e., synthesized UDC notations. These combined notations are used for well-established compound subjects/topics that can only be expressed through a combination of simple UDC notations. At the point of use, in libraries and bibliographic databases the analytico-synthetic nature of UDC generates an infinite number of combinations.

Clearly, some aspects of UDC syntax and relationships could be much better managed if expressed in a more formalized way. Within the semantic web stack (RDF, RDFS, OWL) we can find formal ontology languages with a full apparatus of formal logic that have great power in executing reasoning. Our main focus, however, is not to execute reasoning but to make UDC available as a part of reasoning operations. For this purpose, it is sufficient to express UDC using SKOS and RDFS for the edge labels. At the same time, we have to find ways of expressing some aspects of the UDC syntax, in particular parts dealing with notation synthesis, in a way that is compatible with general web reasoning operations.

For instance, Figure 11 shows how a complex UDC notation 94(492):94(729.885) meaning "The relationship between history of the Netherlands and History of Aruba" can be represented in RDF. To address the problem of the complex number syntax adequately, we deploy two solutions. First, we use "udc-syntax-scheme" to manage different types of notations denoting different types of classes in UDC (main classes, common auxiliaries, special auxiliaries, connecting symbols), as explained in section 4.2. Then we use "blank nodes" to group notational elements belonging to one complex UDC notation. A blank node is a node that has no URI label, just an internal identifier and umbrella pointer to more specific information.

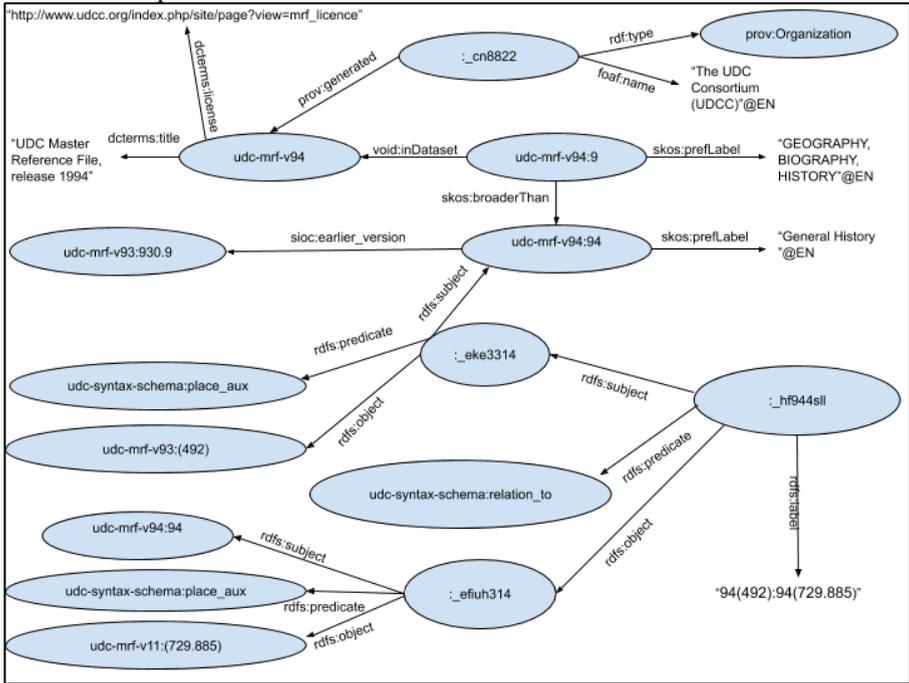

Figure 11. RDF graph representation of a complex UDC notation 94(729.885):94(492).

Our example 94(492):94(729.885) contains the following instances from the UDC scheme:
- a notation from the main table denoting subject 94 General history;
- a connecting sign : (colon) that indicates simple relationships between two subjects (i.e. their notational representation); and,
- two notations enclosed in parentheses which indicate that these are common auxiliaries of place (492) Netherlands and (729.885) Aruba respectively.

In the RDF graph in Figure 11, one can see "udc-syntax-schema:place_aux' as a predicate to the UDC common auxiliary of place (492) Netherlands and as a predicate to the other common auxiliary of place (729.885) Aruba. The syntax predicate 'udc-syntax-schema:relation to'' denotes that two UDC notations, i.e. concepts they represent, are related to each other.xThe blank node ':_hf944sl' indicates that this UDC notation 94(492):94(729.885) is a group that consists of a subject represented by the blank node ":_eke3314," one syntax element "udc-syntax-schema:relation to" and a predicate represented by a blank node ":_efiugh314." The node ":_eke3314" groups notational elements 94 and (492) and node ":_efiugh314" groups notational elements 94 and (729.885).

UDC notational elements in a complex number may originate from different UDC MRF versions. The URI of each notational element indicates the version in which these classes were first introduced. For instance, notation of place (492) was introduced in the version v93, and place (729.885) was first introduced in v11 of the UDC MRF.

In bibliographic collections we expect to find all kinds of new combinations of UDC strings created from simple UDC notations taken from different UDC MRF versions. The method of an "atomic" representation of synthesized UDC notations using the concept of blank nodes, as illustrated above, provides flexibility in supporting analytico-synthetic systems such as UDC.

### 5.3.5 Selection of the RDF schema

As mentioned previously, the UDC look-up service is expected to process requests for URI as well as requests for the full RDF records. Following the parsing stage, URIs for individual notation components and their grouping are generated using RDFs. For the full RDF records we continue to use the SKOS schema extended with several UDC sub-elements. Figure 12 shows the way we mapped the UDC data model (with more specific data elements) to a SKOS schema.

| UDC number (notation) | skos:notation | |
|---|---|---|
| class identifier | skos:Concept | |
| broader class | skos:broader | |
| caption | skos:prefLabel | |
| including note | skos:note | *udc:includingNote* |
| application note | skos:note | *udc:applicationNote* |
| scope note | skos:scopeNote | |
| examples | skos:example | |
| see also reference | skos:related | |
| revision history | skos:historyNote | *udc:revisionHistory* |
| introduction date | skos:historyNote | *udc:introductionDate* |
| cancellation date | skos:historyNote | *udc:cancellationDate* |
| replaced by | skos:historyNote | *udc:replacedBy* |
| last revision data | skos:historyNote | *udc:lastrevisionDate* |

Figure 12. Data elements in UDC LD schema.

The list of 14 data elements contains UDC-specific extensions indicated in italics. Of particular importance for managing historical UDC data, i.e. cancelled classes and their redirection to new classes are *udc:introductionDate*, *udc:cancellationDate* and *udc:replacedBy*.

**6.0 Conclusion**

Publishing the UDC as LD was a research endeavour which served several purposes, some of which are documented in this paper. The UDC case, as we called it within the DiKG project, is relevant beyond the LD publication of this specific KOS alone. The UDC is a representative of a long-lived, widely used KOS in the bibliographic domain. Bringing the UDC (or parts of it) to the LOD cloud entails also mending and establishing links between the design and uses of KOSs prior to the internet and the semantic web technologies available now, which in principle allow deep linking of knowledge and knowledge ordering systems on an unprecedented scale and semantic richness.

The increase in size and the number of information resources (irrespective of the format, language or provenance) and their accompanying KOSs calls for a new generation of approaches. These should allow us to relate (map) and where possible integrate KOSs and their content, with the aim of enabling cross-domain searching and eventually integration of knowledge across different domains on the level of concepts.

Publishing bibliographic classifications as LD has the following potential advantages:

•**Preserve and build on existing classifications data**: Classification notations in resource metadata may be utilised to enable access to the content of a vast number of already classified information resources, i.e., legacy collections in different formats (textual, non-textual, objects), different languages and scripts;

•**Enable navigation and orientation across knowledge spaces from different domains and across different languages and collections:**

•internationally used classification schemes are particularly suitable to be used as a pivot to map different general and special KOSs, thus providing more opportunity for the meaningful linking of collections indexed by different KOSs;

•hierarchical presentation of knowledge fields in bibliographical classifications enables grouping of information on a different level of specificity, and may be used to support information browsing and broadening or narrowing in the information retrieval process and to complement different types of more specific KOSs (e.g. thesauri);

•associative relationships between different knowledge fields and disciplines may be used to enable the presentation of concept dispersion in the knowledge space as a whole and can help in semantic search expansion; and,

•classification schemes translated and containing captions in multiple languages can help in managing connections between notations and language terms for concepts or groups of concepts in many languages; they can help in supporting mapping between classification and thesauri or subject heading systems.

Because the UDC is such an exemplary case for those advantages, we took space in the first sections of this chapter to uncover some of the foundations and related terminology relevant to the understanding of these types of KOSs and their function for information discovery and navigation.

The practical task at hand—publishing the UDC as LOD—was informed by the "10Things Guidelines" (cf. Siebes et al. 2019). As is detailed in the guidelines, each publication process includes conceptual parts (selection of what to publish as LD; how to model this selection as an RDF graph; design of namespaces and URI, etc.) and technological parts (how to ensure machine readability; setting up of the web-based service, etc.). However, as in many LD projects, when applying these principles to the UDC case, we found that the transition to a new technology is never a pure mechanical act. It is a research endeavour in its own right. While our discussions were informed by these guidelines, in the end, not totally unexpectedly, our LD publication project followed its own inner logic. In that some of the generic steps became more important, others vanished into the background, and on the whole the process was much more interwoven and iterative than it appeared in the linear description of the guidelines.

To enable the reader to follow our reasoning concerning the choices we made, we had to first explain the nature of the UDC as existing in the standard UDC data source and those enumerated manifold UDC codes as existing in bibliographic systems around the world. We also discussed four specific challenges which are important to the UDC case that are introduced in Section 4 and (Table 1). The result is the design of an LD service, based on the UDC LD model, which responses in different ways to those four challenges

From the outset, it was clear, that provision had to be made for the inclusion of historical data and for resolution of complex UDC notations, hence the "atomization approach." As discussed in this chapter at various places, general bibliographical classifications are complementary to domain-based KOS design. Their power lies in providing access to concepts (and their various historical and contextual layers) as entailed in the massive number of sources (works) indexed in our collective bibliographical past. Being part of the LOD cloud they have the potential of being used as connectors, similar to manner in which encyclopedias are used, between concepts and their embodiment in works.

| Challenges | Solutions in the LD service design |
|---|---|
| Longevity and system change over time | Inclusion of historical UDC data and concordances between old and ne classes; UDC version-based URI |
| Structural complexity | Provision for expressing UDC syntax in RDF (syntax scheme) and the use blank nodes to allow grouping of notational elements |
| Data ownership | Provision for managing both LOD data set and LD datasets behind the paywall (authentication) |
| Large usage base and amount of legacy data | Provision for parsing, resolving and identifying UDC notations within a look-up service |

Table 1. Challenges and corresponding LD service solutions.

While the design of the LD look-up service is rather practical and modest, in due course, the service provides the potential to augment the UDC namespace which will become better and richer through its use. For example, we can imagine to store resolutions of complex UDC notation queries over time and morph the look-up services to a storage place (or archive) of "all subjects" and topics ever expressed by UDC. In this way, we could provide automatic concordances and conversion from expressions containing obsolete UDC numbers to current UDC expressions.

In creating the design of the new LD service, it also became evident that the LD model is only one part, and that there were other more important tasks. These include preparation of the UDC information (the parsing model) and determining how the LD service relates to the existing functioning technical UDC database structure and editorial system.

Once again, the UDC case has shown the importance of cross-domain, interdisciplinary collaboration which needs experts well situated in the two (or multiple) knowledge domains among which knowledge exchange is supposed to occur. Additionally, we also carried out a lot of work in the background which was needed to care for fruitful knowledge exchange and synergy or the, so-called "trading zone" to use Peter Galison's (1997) metaphor for collaborations in science and technology. Galison was inspired by anthropological studies on collaboration between different cultures in exchanging goods, despite differences in language and culture. This necessitates, from all those involved, a "translatory capability" that enables a cycle of understanding when it comes to each other's conceptual frameworks and terminology, and helps reach a mutual shared understanding which is an indispensable requisite for true interdisciplinary collaboration.

**Notes**
1. Digging into the Knowledge Graph (DiKG) project, 2017-2019 (https://diggingintodata.org/awards/2016/project/digging-knowledge-graph) is funded by the Institute of Museum and Library Services (IMLS) grant within Round 4 of the Digging into Data Challenge grant under the umbrella Trans-Atlantic Platform for the Social Sciences and Humanities. The DiKG research focused on providing means of support for the self-organizing process of knowledge creation in the Semantic Web by enhancing findability and storage for humanities and social science Linked Open Data datasets using the artifacts and organization systems.
2. Bibliographic classifications comprise library classifications (designed for library shelf arrangement) and classifications designed for logical organization and information retrieval of entries in bibliographical services (including abstracting and indexing databases). Library classification are usually less detailed and are structurally simpler than classification designed for use in bibliographies.
3. Classification notations are sometimes called classification codes, classification symbols, classmarks or classification numbers (if notational system is numerical).
4. Classification schemes are not concerned greatly with verbal class descriptions. They differ, in that respect, from thesaurus and subject headings which are primarily concerned with natural language terms used to express concepts in order to manage and control the consistent use of terms. These, for instance, provide alphabetical listing of approved natural language terms (indexing terms) to be used for certain concepts, resolving ambiguities such as homonymy, synonymy or polysemy in a certain field of knowledge but are unable to group or provide logical order of knowledge areas. For this reason, in practice, thesauri and subject headings are usually used as complementary to classifications.
5. The presentation of knowledge space as a whole is a feature of e.g., UDC and Dewey as opposed to Library of Congress *Classification* (LCC) or Bliss Bibliographic Classification (BC2) which function as a series of special classifications.
6. Basic principles on how UDC works are described in numerous books and articles (e.g. McIlwaine 2007). Summary instructions are provided in introduction to all printed UDC editions and an instructional text is provided in UDC Online schedules (www.udc-hub.com).